\documentclass[sn-basic]{sn-jnl}

\usepackage{graphicx}%
\usepackage{multirow}%
\usepackage{amsmath,amssymb,amsfonts}%
\usepackage{amsthm}%
\usepackage{mathrsfs}%
\usepackage[title]{appendix}%
\usepackage{xcolor}%
\usepackage{textcomp}%
\usepackage{manyfoot}%
\usepackage{booktabs}%
\usepackage{algorithm}%
\usepackage{algorithmicx}%
\usepackage{algpseudocode}%
\usepackage{listings}%
\setcitestyle{numbers}
\usepackage{anyfontsize}
\usepackage{graphicx}
\usepackage{subfiles} 
\begin{document}

\title[Article Title]{Physiologically-Informed Predictability of a Teammate's Future Actions   Forecasts Team Performance}

\author*[1]{\fnm{Yinuo} \sur{Qin}}\email{yinuo.qin@columbia.edu}
\author[1,2]{\fnm{Richard} \sur{T. Lee}}\email{rtl2118@columbia.edu}
\author[1]{\fnm{Weijia} \sur{Zhang}}\email{wz2540@columbia.edu}
\author[1]{\fnm{Xiaoxiao} \sur{Sun}}\email{xiaoxiao.sun@columbia.edu}
\author*[1,2,3]{\fnm{Paul} \sur{Sajda}}\email{psajda@columbia.edu}
\affil[1]{\orgdiv{Department of Biomedical Engineering}, \orgname{Columbia University}, \orgaddress{\city{New York}, \state{NY}, \country{USA}}}
\affil[2]{\orgdiv{Department of Electrical Engineering}, \orgname{Columbia University}, \orgaddress{\city{New York}, \state{NY}, \country{USA}}}
\affil[3]{\orgdiv{Department of Radiology}, \orgname{Columbia University}, \orgaddress{\city{New York}, \state{NY}, \country{USA}}}


\abstract{In collaborative environments, a deep understanding of multi-human teaming dynamics is essential for optimizing performance. However, the relationship between individuals’ behavioral and physiological markers and their combined influence on overall team performance remains poorly understood. To explore this, we designed a triadic human collaborative sensorimotor task in virtual reality (VR) and introduced a novel predictability metric to examine team dynamics and performance. Our findings reveal a strong connection between team performance and the predictability of a team member's future actions based on other team members' behavioral and physiological data. Contrary to conventional wisdom that high-performing teams are highly synchronized, our results suggest that physiological and behavioral synchronizations among team members have a limited correlation with team performance. These insights provide a new quantitative framework for understanding multi-human teaming, paving the way for deeper insights into team dynamics and performance.}




\maketitle
\section{Introduction}\label{intro}
Teamwork is a critical form of human interaction, productivity, and survival. From world-championship sports teams to intimate working groups, from ancient tribal rituals to modern urban planning, teaming has consistently been a critical and innate element of human behavior. Without teaming, our society would likely look very different, lack rich and diverse cultures, lack marvels of engineering construction, and have limited groundbreaking scientific advancements. Studying the fundamental mechanisms behind human teaming is essential to understanding and improving collective human intelligence. 

Games and collaborative tasks have been used as major platforms to study multi-human teaming. From role-playing to battle arena games, many previous studies have shown that multiplayer online games have great potential for studying team dynamics \cite{dabbish2012communication}, leadership in multi-human teaming \cite{jang2011exploring, pobiedina2013successful}, and individual behavior within teams \cite{sapienza2018individual}. However, it is still unclear whether the insights gained from simple game-based studies can be generalized to more complex, high-stakes team interactions and team performance.

In addition to computer games, real-world scenarios, such as simulated hospitals with surgical teams and teams in manufacturing companies, have been used to study team performance and effectiveness \cite{morgan2007evaluating, edmondson1999psychological}. Most of these previous studies have used qualitative methods such as interviews \cite{edmondson1999psychological}, questionnaires \cite{edmondson1999psychological, pearsall2006effects}, and surveys \cite{morgan2007evaluating, van2008critical}. While these qualitative studies can help us gain insights into how some task-related factors can impact team performance, these methods are prone to bias due to subjective reporting and are often difficult to reproduce. Therefore, additional consideration of quantitative evaluation metrics to understand team performance remains essential but largely unexplored.

With the development of virtual reality (VR), more environmentally controlled team-based studies have been conducted \cite{varlet2013virtual, hansen2020asymmetrical, moore2020familiarity, weissker2021group}. VR provides an immersive experience while reducing external distractions. The virtual environment also has the potential to provide realistic simulations with well-controlled delivery and simultaneous recording of events and interactions. However, few VR experiments have involved real-time synchronization and multi-modal data collection of multi-person teams. Most team-based VR experiments are conducted with a single human performing collaborative tasks with other simulated computer agents instead of working in the simulation with other people \cite{hansen2020asymmetrical, moore2020familiarity}. These experiments limit the possibility of studying multi-human teaming.

Through studying human teaming in various tasks, previous research has highlighted that physiological synchrony among team members is positively correlated with team performance \cite{henning2001social, dikker2017brain, gordon2020physiological, madsen2022cognitive}. Conversely, other studies have suggested a negative correlation between behavioral synchrony and team performance \cite{abney2015movement, vicaria2016meta}. The preceding literature lacks studies that comprehensively correlate performance with both behavioral and physiological synchrony in complex teaming tasks. The interpretation of such correlations of team performance with behavioral synchrony and physiological synchrony remains unclear and incomplete. Therefore, we hypothesized that a comprehensive understanding of the balance between physiological and behavioral synchrony is critical for enhancing team performance, especially in tasks that demand high levels of cooperation, coordination, and collaboration. 

In this work, we developed a novel framework to study multi-human teaming in a VR environment by quantitatively analyzing multi-modal physiological and behavioral data from all team members. We constructed an immersive sensorimotor task requiring three participants to collaboratively navigate a spacecraft, capturing multi-modal data from all participants. To identify potential biomarkers of team performance, we employed two computational approaches: inter-subject correlation (ISC) and predictability. ISC, traditionally linked to team performance metrics \cite{szymanski2017teams, reinero2021inter}, was found to correlate with team performance only under specific measurements in our complex collaborative task. To address these limitations, we proposed a predictability approach, using a deep learning model to forecast one team member’s remote controller actions based on their teammates' physiological and behavioral data. This predictive model revealed a significant correlation between the predictability of team members’ actions and team performance, suggesting that predictability can serve as a robust biomarker for understanding and enhancing team dynamics in collaborative tasks.

\section{Results}\label{results}
\begin{figure}[htp]
    \centering
    \includegraphics[width=1\textwidth]{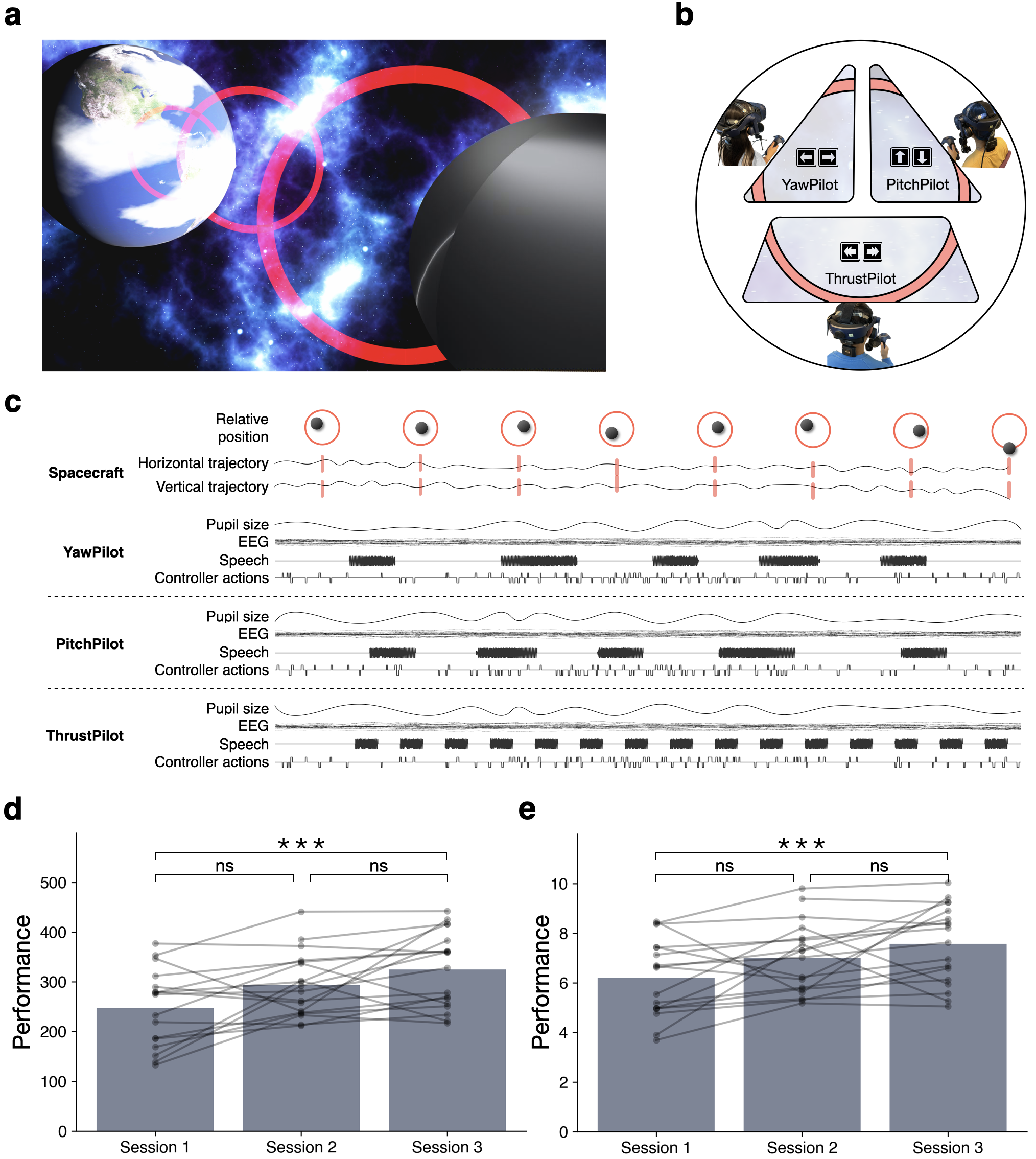}
    \caption{ADCT environment and team performance. \textbf{a}, An illustration of ADCT virtual environment. The team's goal is to control the spacecraft, passing all red rings and arrive back to Earth within the specified time limit. \textbf{b}, The view of three co-pilots with respect to a ring obstacle and the degree of freedom controlled by each role. The three co-pilots are YawPilot, PitchPilot, and ThrustPilot. Each participant was equipped with a VR headset, a microphone, a remote controller, and an EEG headset. \textbf{c}, Illustration of data modalities collected from all co-pilots. The red bars on the spacecraft's horizontal and vertical trajectories represent the relative location of ring obstacles. The uppermost section illustrates the cross-section of a spacecraft's position with respect to a ring. \textbf{d-e}, Team performance across three experimental sessions. The performance is measured by the number of rings passed. Each dot indicates one team (N = 17). Bars indicate the average across teams. \textbf{d}, Total number of rings passed by each team in each session. \textbf{e}, Averaged the number of rings passed by each team in each trial in three sessions. Asterisks indicate statistically significant differences, defined as ns, not significant, $***P < 0.001.$}
    \label{fig:experiment_plt_1}
\end{figure}
To test the correlation between team performance and physiological and behavioral synchrony among team members, we designed a multi-human team-based virtual reality (VR) task that we refer to as the \textbf{A}pollo \textbf{D}istributed \textbf{C}ontrol \textbf{T}ask (ADCT). Our task is inspired by the renowned Apollo 13 reentry mission and its extended cinematic story \cite{rerup2001houston, Apollo13_1995}. The Apollo 13 mission is considered one of history's most ``successful failures'' in that three astronauts exhibited extraordinary teamwork while operating different controls of a spacecraft collaboratively to safely navigate back to Earth after an oxygen tank exploded. The ADCT is a team-based version of a boundary avoidance task (BAT) \cite{Faller:2019dy}, which requires substantial attention and regulation of arousal of each individual in the team. The ADCT has the following features built into its design and construction: 1) it is a challenging enough cooperative and collaborative task to trigger complex team dynamics; 2) the experiment was conducted repetitively with a consistent group of participants; 3) the task state and behavior of subjects are synchronized in real-time with simultaneously recorded multi-modal physiological signals; 4) team performance is quantitatively assessed by evaluating the contributions of all team members, where local performance is measured in relation to short-term goals, and global performance encompasses high-level planning tragedies. Fig. \ref{fig:experiment_plt_1} summarizes the ADCT. 

Specifically, the ADCT is performed by a triad team in VR (Fig.\ref{fig:experiment_plt_1} a). Each team member, as a co-pilot, has partial observation of the exterior space environment through uniquely positioned spacecraft windows, each with different viewing points. Each co-pilot controls a single degree of freedom of the spacecraft's movement, such as yaw, pitch, or thrust (Fig.\ref{fig:experiment_plt_1} b). The team's goal is to collaboratively navigate the spacecraft back to Earth by following a pre-defined reentry path. The transparent red rings mark the boundary of the path, and the team must reach Earth within a limited time. Therefore, failing to pass all rings with sufficient speed results in trial failure. Teams are monetarily incentivized to complete as many trials successfully as possible. If they cannot return to Earth in time, they must navigate the entry path to get as close to Earth as possible.

While the teams performed the ADCT, we simultaneously collected electroencephalography (EEG), pupillometry, eye gaze, speech, and remote controller inputs from all participants (Fig.\ref{fig:experiment_plt_1} c). Each team participated in three experimental sessions. The roles of participants were randomly assigned for each session, but the team members remained the same across all sessions. Each experimental session included 45 trials, each consisting of 15 rings. Team performance was quantitatively evaluated by the team's total number of ring obstacles successfully navigated.

\subsection{Team Performance Improves Across Experimental Sessions} 
We first analyzed performance dynamics across three experimental sessions to investigate how physiological and behavioral synchrony among team members relates to team performance. As shown in Fig.\ref{fig:experiment_plt_1} d, the total number of rings passed by each team increased monotonically over the experimental sessions, indicating a steady improvement in overall team performance. Repeated measures analyses of variance (ANOVA) revealed significant performance differences across sessions ($F(2,32) = 10.88,~ p < 0.001,$). Post-hoc comparisons with Bonferroni correction showed a substantial improvement in performance from Session 1 to Session 3 ($p < 0.001$). Similarly, the averaged trial performance also improved significantly over time (Fig.~\ref{fig:experiment_plt_1} e, $F(2,32) = 7.75,~ p < 0.01$). The performance significantly increased from Session 1 to Session 3 ($p < 0.001$). These findings suggest that team performance improved consistently as participants engaged in more task sessions. This steady enhancement highlights the potential for learning and adaptation in team dynamics through repeated collaborative tasks in immersive environments.

\begin{figure}[htp]
    \centering
    \includegraphics[width=1\textwidth]{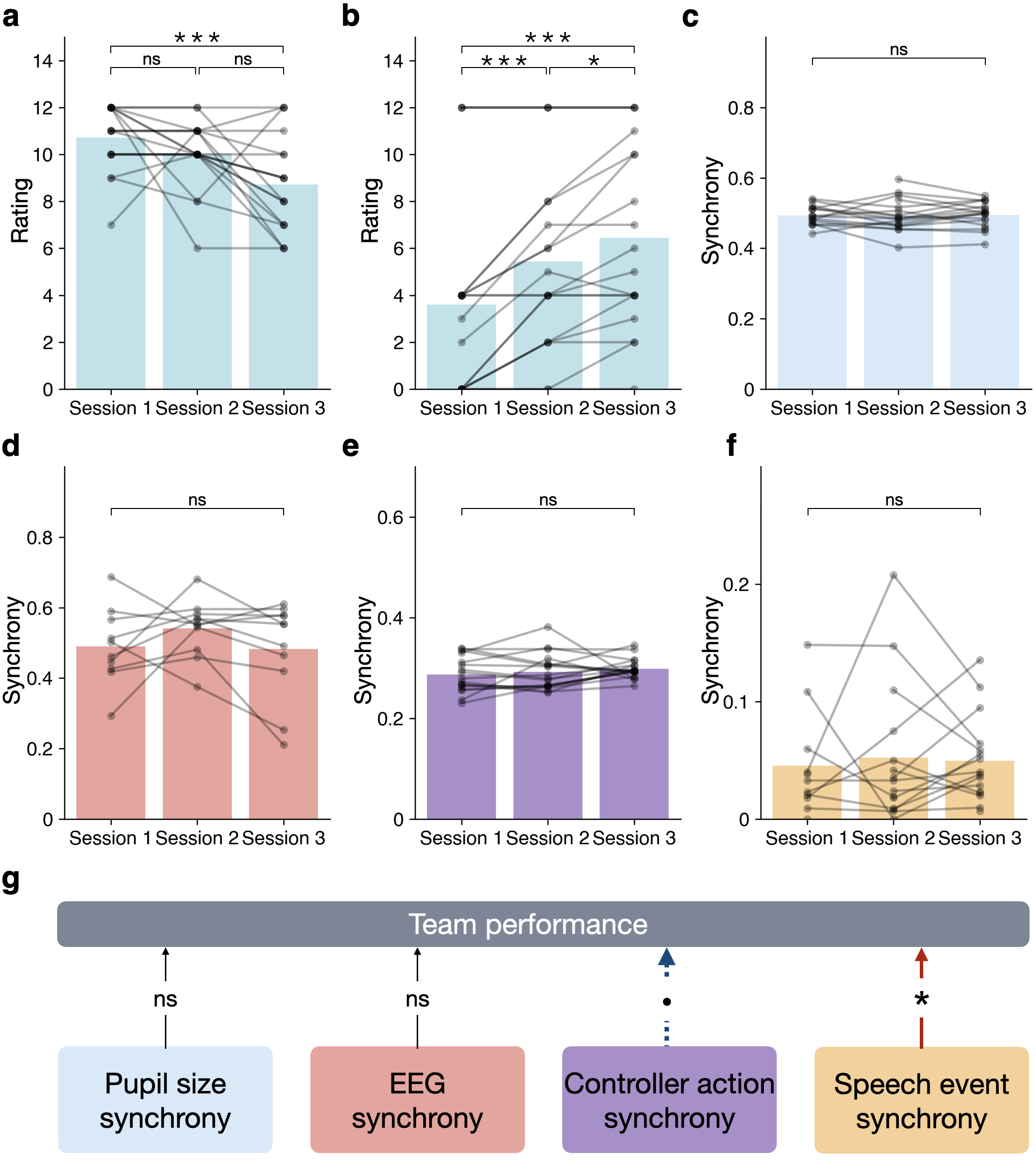}
    \caption{Subjective rating and multi-modal synchrony. \textbf{a-f}, Subjective ratings and synchrony among team members based on different physiological or behavioral data modalities across experimental sessions. Each dot represents one team, and the bars show the average across teams. \textbf{a}, Helpfulness rating of team members (N=17).  \textbf{b}, Familiarity rating of team members (N=17). \textbf{c}, Pupil size synchrony among team members (N=14). \textbf{d}, EEG synchrony among team members (N=9). \textbf{e}, Remote controller action synchrony among team members (N=17). \textbf{f}, Speech event synchrony among team members (N=7). \textbf{g}, Multi-modal synchrony and its correlation with team performance. Blue arrows indicate negative correlations, while red arrows indicate positive correlations. Asterisks indicate statistically significant differences, defined as ns, not significant, $\boldsymbol{\cdot} P<0.1, *P < 0.05, ***P < 0.001.$}
    \label{fig:performance_familiarity_synchrony}
\end{figure}

\subsection{Subjective Ratings and Multi-Modal Inter-Subject Synchrony} 
After each experimental session, all co-pilots provided subjective ratings of their familiarity with and helpfulness toward other team members (see Post-Task Survey in \nameref{post_task_survey} for details). Analyzing these ratings allows us to track how familiarity and helpfulness change over time and investigate the potential impact of team members' perceptions on team performance. Surprisingly, the helpfulness rating shows a consistent decrease across the experimental sessions (Fig. \ref{fig:performance_familiarity_synchrony} a, repeated measures ANOVA, $F(2,34) = 9.33, p<0.001$). In contrast to the decreasing helpfulness scores, the average familiarity rating across teams increases monotonically (Fig.\ref{fig:performance_familiarity_synchrony} b, $F(2,34) = 21.42, p<0.001$). This pattern suggests that as team members become more familiar with each other, their perceptions of helpfulness may become more critical or nuanced.

Next, we analyzed the dynamics of team synchronization by calculating the inter-subject correlation (ISC) across various data modalities. ISC is a widely recognized metric for evaluating the synchrony among individuals performing identical tasks \cite{kauppi2010inter,dmochowski2012correlated, dikker2017brain, poulsen2017eeg, keles2024multimodal} or collaborative tasks \cite{vspilakova2020getting, xie2020finding}. This work analyzed the ISC among three co-pilots based on their pupil dynamics, EEG, remote controller inputs, and speech events. As illustrated in Fig.\ref{fig:performance_familiarity_synchrony} c, pupil size synchrony remains relatively stable across sessions ($F(2,28) = 0.65, p=0.53$). Interestingly, EEG ISC is maximized in the second experimental session. However, variations in EEG ISC did not achieve statistical significance (Fig.\ref{fig:performance_familiarity_synchrony} d, $F(2,18) = 1.51, p=0.25$). Remote controller actions and speech events also remain stable along the three experimental sessions (Fig.\ref{fig:performance_familiarity_synchrony} e, f; remote controller actions synchrony $F(2,32) = 1.10, p=0.34$; speech event synchrony $F(2,14) = 0.23, p = 0.80$). These findings suggest that increasing experimental sessions have a limited impact on synchronizations among team members' behavioral or physiological data.

\subsection{Inter-Subject Synchrony and Its Correlation with Team Performance}
Inter-subject synchrony (ISC) is often hypothesized to be correlated with team performance. Previous studies have demonstrated a positive relationship between team performance and synchrony in brain and pupil dynamics \cite{szymanski2017teams, dikker2017brain, reinero2021inter, wohltjen2023synchrony}. However, whether synchrony among more than two team members correlates with overall team performance remains unexplored. To address this, we employed generalized linear mixed-effects models (GLMMs) to examine the relationship between inter-subject synchrony across multiple modalities and team performance (see \nameref{GLMM} for details).

Our findings reveal that behavioral synchrony, such as controller action synchrony and speech event synchrony, significantly correlate with team performance (Fig. \ref{fig:performance_familiarity_synchrony} g). Interestingly, speech event synchrony among team members is positively correlated with team performance, suggesting that verbal coordination enhances high-level task outcomes ($\beta = 1.63, P = 0.039$). In contrast, controller action synchrony is negatively correlated with team performance, possibly reflecting a detrimental effect of over-coordination on individual autonomy in control actions ($\beta = -1.01, P = 0.072$). Physiological synchrony, however, did not show a significant correlation with team performance (pupil size synchrony, $\beta = -0.73, P = 0.328$; EEG synchrony, $\beta = 0.11, P = 0.845$). These results show that behavioral synchrony is a key predictor of team performance in triad teams, highlighting a previously overlooked factor in team performance research.

\begin{figure}[htp]
    \centering
    \includegraphics[width=0.8\textwidth]{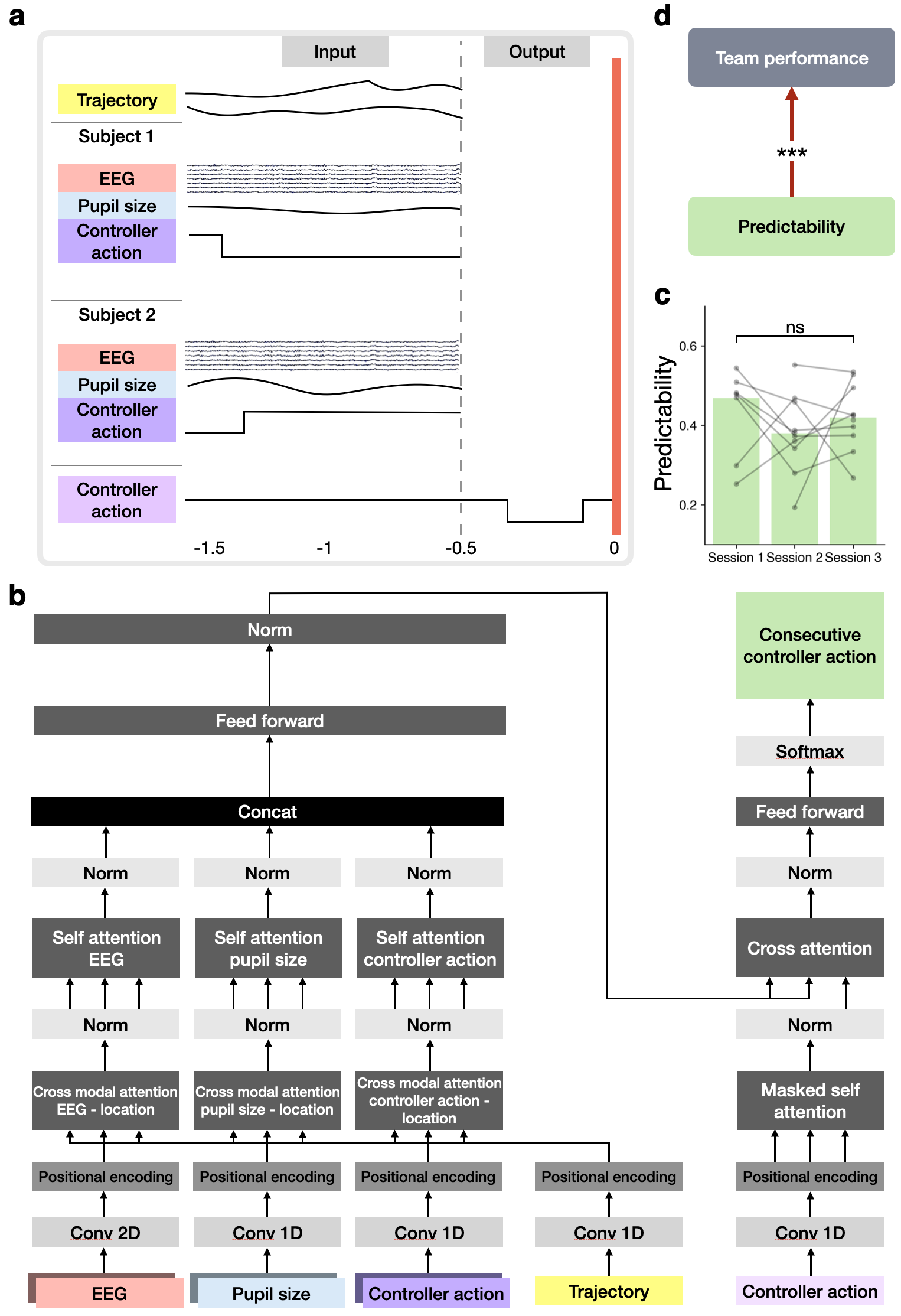}
    \caption{Predictability of each team member's actions as a biomarker of team performance. \textbf{a}, An illustration of a single epoch of the multi-modal data. Each epoch is relative to a ring, and we divided each epoch into input and output for the predictive model. The predicted action of an individual is based on a generative model that uses the behavioral and physiological data of the other two co-pilots. Predictability is evaluated by computing the correlation of the true action of a co-pilot with the model-predicted action. \textbf{b}, Multi-head attention modal structure. The cross-modal attention layers take the spacecraft trajectories and physiological or behavioral data. \textbf{c} Team predictability across three experimental sessions. Each dot represents one team, and the bars show the average across teams ($n=10$). ns, not significant. \textbf{d}, Correlation between team performance and predictability. The red arrow indicates positive correlations. Asterisks indicate statistically significant differences, defined as $***P < 0.001.$}
    \label{fig:predictability}
\end{figure}

\subsection{Quantifying Team Predictability Using Multi-Modal Physiological and Behavioral Data}
A high-performing team consists of members who consistently engage in predictable interactions \cite{bradley2013team}. This predictability results from a deep understanding and harmony within the team, making it easier for team members to anticipate one another's actions and reactions to each other. In this study, we used a multi-head attention model to quantitatively measure how the future actions of a teammate could be predicted from their teammates' physiology and behavior. 

First, we epoched multi-modal physiological and behavioral data from 1.5 seconds before each ring-passing event (Fig.\ref{fig:predictability} a). The model received inputs from the initial 1 second of this epoch, where each input included the spacecraft's trajectory and the behavioral and physiological data of two co-pilots. The model's output was the generated prediction of the constructive 0.5-second controller action of the third co-pilot (0.5 seconds before passing the ring). On average, co-pilots made about 0.3 remote controller actions in that time period. We evaluated predictability at the team level by averaging the individual predictability scores across the three co-pilots. Since speech event synchrony significantly correlates with team performance ($P < 0.05$), we excluded speech event data from the model input to avoid potential bias. (The supplementary materials include results from a model incorporating speech input for comparison.) By analyzing team predictability, we demonstrate its potential as a biomarker significantly associated with overall team performance.

This model architecture addresses the challenge of integrating multi-modal data with varying temporal and spatial characteristics (Fig.\ref{fig:predictability} b). Including cross-modal attention layers ensures that inter-modal dependencies are captured, particularly when aligning trajectory information across diverse behavioral and physiological data modalities. The self-attention layers for each modality are crucial for extracting meaningful intra-modal patterns, such as EEG synchrony or patterns in pupil size dynamics. By concatenating the outputs of all modalities, the feed-forward network integrates complementary features, creating a unified representation that captures the interactions between physiological, behavioral, and environmental data.

The cross-attention mechanism bridges the fused multi-modal representation with the target modality, facilitating accurate predictions of controller actions. This structure allows the model to leverage the unique contributions of each modality while ensuring robustness in handling noisy data. The design promotes modularity and adaptability, making it suitable for analyzing multi-modal tasks requiring temporal and spatial alignment, such as collaborative team performance or dynamic decision-making tasks.

We hypothesized that the predictability of team members' future controller actions would significantly correlate with team performance. Consequently, we expected that the predictability of each team’s actions would change across experimental sessions. As shown in Fig.~\ref{fig:predictability} c, predictability changes slightly as the number of experimental sessions increases ($F(2,10) = 0.12, P=0.888$). A detailed analysis of predictability and team performance is provided in the next section.

\subsection{Team Action Predictability as a Performance Biomarker}
We derive an intriguing finding that predictability serves as an important biomarker for team performance (Fig. \ref{fig:predictability} d, $\beta = 3.20, P<0.001$). The positive correlation with team performance suggests that when team members can better anticipate each other’s future actions, overall coordination improves, enabling the team to achieve higher-level goals such as passing more rings. 

\begin{figure}[htp]
    \centering
    \includegraphics[width=0.8\textwidth]{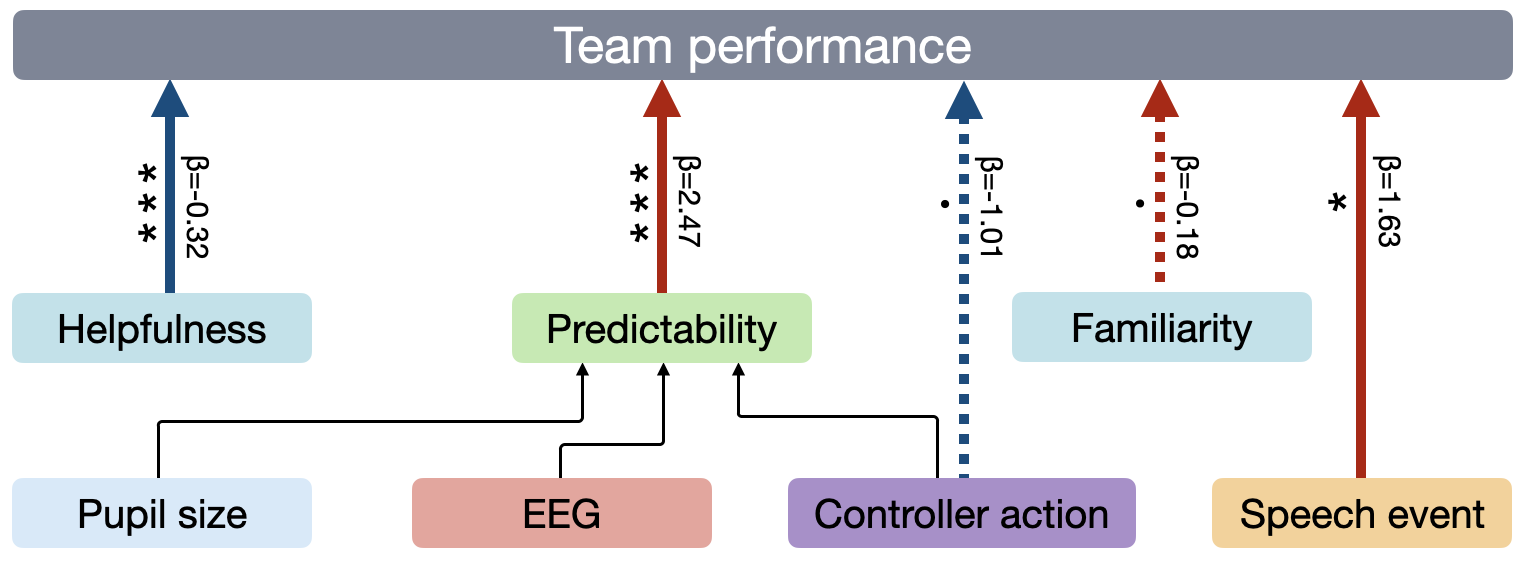}
    \caption{Overview of the correlation between predictability and team performance. All correlations are the GLMM results in accounting for session-level differences. The predictability biomarker is computed based on multi-modal physiological and behavioral data. Each arrow originates from the independent variable and points towards the dependent variable. The blue arrow indicates negative correlations, while the red arrow indicates positive correlations. Dashed lines indicate insignificant correlations. $\boldsymbol{\cdot}~P < 0.1, * P < 0.05, *** P < 0.001$, n = 10. }
\label{fig:performance_predictability_synchrony}
\end{figure}

\subsection{Team Performance and Subjective Ratings of Team Members}
Familiarity among team members has been demonstrated to be positively correlated with team performance \cite{harrison2003time, gordon2020physiological}. Our experiment focused on a collaborative distributed control task observed a similar pattern (Fig. \ref{fig:performance_predictability_synchrony}). Specifically, familiarity among co-pilots is positively correlated with team performance, indicating that as familiarity increases, teams perform better on both sub-tasks and in achieving their long-term goals ($\beta = 0.18, P=0.074$). This finding emphasizes the importance of building familiarity within teams, as it appears to enhance their ability to work cohesively and effectively toward shared goals.

Interestingly, the helpfulness rating of team members is significantly negatively correlated with team performance ($\beta = -0.32, P<0.001$). This suggests that higher helpfulness ratings may reflect a greater reliance on teammates for support, which could reduce individual autonomy or efficiency, potentially detracting from the overall team performance. Conversely, lower helpfulness ratings may indicate a more balanced contribution from all members, optimizing team efficiency toward achieving shared objectives. Together, the subjective ratings of familiarity and helpfulness reveal a nuanced relationship between team dynamics and performance. While familiarity fosters cohesion and shared understanding, perceptions of helpfulness may introduce dynamics that negatively impact team performance. These insights highlight the complex interplay between subjective perceptions and performance, offering valuable guidance for designing and optimizing collaborative teams in distributed control tasks.

\section{Discussion}
In this study, we conducted a team-based collaborative virtual reality (VR) experiment and demonstrated a novel multi-modal biomarker that directly correlates with team performance. Specifically, we demonstrated that a biomarker measuring the predictability of teammate behavior is better correlated with team performance. This biomarker is derived from integrating multi-modal physiology and behavior of teammates to predict the future behavior of the remaining (i.e. left out) team member. Our predictability biomarker challenges the conventional wisdom that physiological and behavioral synchrony is a robust marker of a high-performing team \cite{henning2001social, abney2015movement, vicaria2016meta, gordon2020physiological, madsen2022cognitive}. 

Simultaneously collecting and analyzing multi-modal data is crucial for understanding team performance and dynamics. In contrast to executing simple tasks individually, collaborative tasks involve complex dynamics and interactions among team members. Various data modalities, including pupillometry, EEG, speech, and other physiological or behavioral data, have been analyzed individually but not in combination \cite{doi:10.1177/2327857916051010, mccraty2017new, dias2019physiological, dunbar2020using}. We have developed a cross-modal multi-head attention predictive model that is capable of simultaneously analyzing multi-modal data from multiple team members (Fig.~\ref{fig:predictability} b). This model integrates inputs from multiple data modalities, enabling not only the prediction of future actions of individuals but also the identification of a biomarker that is inversely related to overall team performance \cite{madsen2023concept}. This result further demonstrates that different physiological and behavioral measures provide unique information that needs to be integrated to construct biomarkers that better relate to performance. 

Our results revealed a positive correlation between our predictability biomarker and team performance. Aligning with the common belief that high-performance teams benefit from predictable actions among members \cite{bradley2013team}, our findings suggest that this is expresses in teammate physiology in a way that leads to enhanced coordination and alignment for achieving higher performance (Fig.~\ref{fig:performance_predictability_synchrony}). While the synchrony of individual modalities among co-pilots showed marginal or insignificant correlation with team performance, combining multi-modal data as input to the predictive model revealed that the predictability of team members’ future actions is a stable and reliable biomarker of team performance. This highlights the potential of leveraging predictability as a key metric for understanding and improving team dynamics.

We have focused primarily on using predictability as a key indicator of team performance in collaborative tasks involving multiple humans. A pivotal question arising from our research is how we may practically leverage the predictive abilities of team members to enhance team dynamics and performance. This capability can facilitate collaboration between humans or, potentially, teaming between humans and artificial intelligence (AI) agents \cite{chen2021visual, metcalfe2021systemic, harris2023social}. Our findings lay the groundwork for innovative teaming strategies, fostering enriched and more productive collaborations.

\section{Methods}\label{methods}
\subsection{Participants}
Fifty-four healthy human participants (age $= 23.67 \pm 3.34$ year (mean $\pm$ standard deviation); 27 females, 27 males) voluntarily participated in the three experiments. These participants were divided into 18 triad teams, and each team participated in three sessions on different days. Due to incomplete sessions, data from one team were omitted from the final data set. Data from four teams were omitted from the pupil size analysis due to invalid pupil size recordings of one or more co-pilots. EEG data from nine teams were excluded from the analysis due to error-prone recordings identified during preprocessing. Similarly, speech event data from ten teams were excluded because the speech event detection algorithm failed to extract speech events from one or more participants within the team. No participants or teams dropped out of the experiment due to motion sickness or other symptoms related to virtual reality. All participants had normal or corrected-to-normal visual acuity and gave informed consent before participating in each experiment. Human subject protocols were approved by the Columbia University Institutional Review Board. 

\subsection{Virtual Environment}
The virtual environment was built using \textit{Unreal Engine 4.25.4}. The four main reactive objects in the virtual environment were 1. a spacecraft, 2. a countdown timer, 3. the rings, and 4. the Earth. As shown in Fig.\ref{fig:experiment_plt_1} b, three viewing windows with different shapes and at different positions were placed at the front of the spacecraft. Each subject in the triad team was assigned to look through one window, and the degree of freedom the subject controls was fixed per experiment session, corresponding to its respective window. The ThrustPilot, who controlled the speed of the spacecraft, had the largest unobstructed field of view, which was located at the bottom of the spacecraft. The YawPilot, who controlled the left-right spacecraft movement, was located at the top-left of the spacecraft. The PitchPilot had a viewing window on the top-right and controlled the up-down movement of the spacecraft. Because the positions and shapes of the windows were different, subjects with different roles had partial and biased views of the environment. The field of view of the virtual camera of each co-pilot is 80\textdegree in Unreal Engine. 

A countdown timer bar was displayed at the bottom of each window to indicate the remaining time for each trial. Initially, the timer bar was completely black. As time elapsed, the black portion of the bar gradually decreased, revealing an increasing white segment. This white segment represented the time that had passed and was inversely proportionate to the black portion, which showed the remaining time. Each trial had a maximum duration of 55 seconds. The timer would automatically stop and reset if the team either successfully navigated through all the rings and approached Earth, or failed to pass through any ring during the trial. Despite this, with the default speed of the spacecraft, teams would require at least 60 seconds to pass through all rings in a trial, presenting a significant challenge and requirement for active participation and collaboration with the ThrustPilot.

The rings were transparent red toruses that represented a trial's reentry path. At the beginning of each trial, a sequence of fifteen rings was generated, spaced equally but positioned at varying horizontal and vertical coordinates. The distance between any two adjacent rings was $50,000$ units in \textit{Unreal Engine}. The Earth was positioned at the end of the path with $50,000$ units from the final ring. The trial ended when the spacecraft, operated by the team of participants, successfully navigated through all rings and stopped in front of Earth. Upon successful completion, the term ``Successful" will be displayed on each participant’s head-mounted display (HMD) for one second. Subsequently, a new trial will automatically be started. 

\subsection{Apparatus}
In all experiments, each participant was equipped with VIVE Pro Eye head-mounted displays (HTC Corporation; resolution: $1440 \times 1600$ pixels per eye; refresh rate: 90 Hz), and an EEG device with 20 electrodes was placed in accordance with the international 10-20 system (Advanced Brain Monitoring B-Alert X24; sample rate: 256 Hz). A USB microphone was set in front of each subject to enable communication between subjects, and Mumble (version 1.4.230) was running locally on each desktop. We used LabRecorder (version 1.14.0) to collect the multi-modal data. Each head-mounted display is connected to a desktop with an Intel Core i9 CPU and an NVIDIA RTX 2070 Super GPU. The three desktops were connected to a local, secure WiFi network with a 2.6 Gbps router using client-server network protocols to communicate. The server was another desktop with an Intel Core i9 CPU and an NVIDIA RTX 2080 Super GPU. 

\subsection{Procedure}
In each experiment, three participants arrived at the lab and watched an instructional video before the first session. Following the setup of the EEG devices, participants were escorted to three separate EEG recording chambers designed to block sound and electrical noise. These chambers were additionally acoustically shielded with 2-inch thick soundproofing foam to prevent echoes and minimize noise interference. We assisted the participants in setting up head-mounted displays and remote controllers.

Individual eye calibration commenced once each participant was fully equipped and settled. The calibration was conducted using the VIVE Pro Eye system. Each experiment began with five pilot trials following eye calibration, allowing subjects to familiarize themselves with the task environment before the commencement of data collection. A trial was terminated when the team failed to pass a ring due to a crash or a miss or if the time limit was exceeded. After the pilot trials, participants were notified via headphones that the experiment had officially started.

Each team participated in three repeated sessions of the same experiment. Each session was spaced at least 24 hours apart, and no participant had participated in nor had familiarity with the task before their first session. Within each experiment session, roles were randomly assigned to the subjects. After each experimental session, all participants were asked to complete a post-task questionnaire separately (see the post-task survey in \nameref{post_task_survey} for details). 

\subsection{Data Preprocessing}
We implemented different pre-processing methods for various data modalities. For pupil size data, we first detected and removed blinks and artifacts. Then, we applied linear interpolation and Z-scored the pupil size data. This was followed by averaging the pupil size between the left and right eyes and a fourth-order Butterworth lowpass filter to remove high-frequency noise. 

The EEG pre-processing included filtering the raw EEG data using fourth-order Butterworth bandpass filters with bands 0.5 Hz-100 Hz (MNE 1.6.1 \cite{GramfortEtAl2013a}). Manual bad channel rejection was conducted to remove error-prone channels in each recording. Then we performed Independent Component Analysis (ICA) \cite{makeig1995independent} and used the Multiple Artifact Rejection Algorithm (MARA) \cite{winkler2011automatic} to separate and reject artifact components. 

Remote controller actions were first down-sampled to 60 Hz. Next, values greater than 0.5 were assigned a value of 1, values less than -0.5 were assigned a value of -1, and values between -0.5 and 0.5 were assigned a value of 0. 

The speech preprocessing involved three steps. First, we applied the noise reduction function \cite{sainburg2020finding} to the speech recordings from each subject to remove background noise. Next, we used a simple voice activity detection function to extract speech events. Finally, the speech events were down-sampled to 60 Hz. All data modalities were then epoched based on the relative time to the respective rings and saved for analysis.
\subsection{Post Task Survey} \phantomsection\label{post_task_survey} 
After each experimental session, all participants were asked to complete a survey comprised of demographic and subjective questions. In this study, our analysis concentrated on two specific subjective questions:
\begin{enumerate}
    \item \label{Q1} How helpful was each of your teammates in reaching the final goal?
    \item \label{Q2} How well did you know each of your team members before today?
\end{enumerate}
Each participant was required to select one of three possible answers for each question that concerned every other team member, excluding themselves. These answers were scaled as 0 = Not at all, 1 = A little, and 2 = Very well. The responses to the helpfulness (Question \ref{Q1}) and familiarity (Question \ref{Q2}) questions were assessed based on the team and the specific experiment session. The helpfulness and familiarity scores ranged from 0 to 12 for each team. A score of 0 indicated that all three participants rated `Not at all' for each of the other two team members. In contrast, a score of 12 indicated that every participant rated `Very well' for their teammates.

\subsection{Pupil Size, Remote Controller Action, and Speech Event Synchronies}
This study computed the inter-subject correlation (ISC) across the three subjects using their pupil sizes, remote controller actions, and speech events. For each experiment session, we computed the Pearson Correlation Coefficient ($r$) between each pair of participants, participant $a$ and participant $b$, with their distinct roles within the same team, for one data modality at a time, using (\ref{p_correlation_1})
\begin{align}\label{p_correlation_1}
    r_{a, b} &= \frac{\sum _{i = 1}^n (a_i - \overline{a}) (b_i - \overline{b})}{\sqrt{\sum _{i = 1}^n (a_i - \overline{a})^2\sum _{i = 1}^n (b_i - \overline{b})^2}}.
\end{align}

$n$ was the length of one epoch of data. The team ISC in one epoch $r_{epoch}$ was the averaged ISC across three co-pilots. 

\subsection{EEG ISC}
To assess inter-brain synchronization, we computed ISC using Correlation Component Analysis (CorrCA) \cite{parra2018correlated}. This approach involved utilizing linear combinations of EEG channels or EEG signals with other data modalities as separate channels that maximize the ISC on the data obtained from subjects within the same team. In our study, we employed an improved version of CorrCA to compute the correlation between multiple subjects within the same team while performing a collaborative control task. The EEG signals of each subject contained $20$ channels, and the approach finds a weight vector $w$ that maximizes the Pearson Correlation between subjects in the team. 

The weight vector $w$ determines which linear combination of different channels provided the most significant correlation among team members. Given the EEG signals of the three subjects, denoted as $X_1, X_2$, and $X_3$, where $X_n \in \mathbb{R}^{D\times T}$ with $D$ representing the number of channels and $T$ representing the number of time steps in an epoch, the weight vector $w$ could be computed by:
\begin{align}
\begin{split}
    &w = arg max_w(\frac{w^TR_{12}w}{\sqrt{w^TR_{11}w}\sqrt{w^TR_{22}w}});\\
    &\text{where } R_{ij} = \frac{1}{T}X_iX_j^T  
\end{split}
\end{align}
We defined the within subject covariance as $R_w = \sum_i^N R_{ii}$ and between subject covariance as $R_b = \sum_i^N \sum_{j, j>i}^N R_{ij}$. 
Here, $N=3$ denoted the number of subjects in each experiment. We computed the eigenvectors $e_k$ of $R_w^{-1}R_b$ and ranked the eigenvectors in descending order based on the corresponding eigenvalues. Hence, the ISC was the maximum value of the strengths of correlations $C_k$, where
\begin{align}
    C_k = \frac{e_k^T R_b e_k}{e_k^T R_w e_k}.
\end{align}

\subsection{The Generative Forecasting Model} 
The predictive model we implemented was a multi-head attention-based neural network that tracked relationships between events in data within the time domain. Fig. \ref{fig:predictability} b illustrates the structure of the model. The input to the model included the team's spacecraft trajectory along with the behavioral and physiological data of two participants. The transformer model utilized both encoders and decoders discussed in the original transformer model \cite{vaswani2017attention}. The 8-head attention layers in the encoder and the masked 8-head attention layers in the decoder were implemented as follows:
\begin{align}
\begin{split}
    & MultiHead(Q, K, V) = Concat(head_1, ..., head_4)W^O,\\
    & \text{where } head_n = Attention(QW^Q_n, KW^K_n, VW^V_n),\\
    & [W^Q_n, W^K_n] \in \mathbb{R} ^{d_{m} \times d_k}, W^V_i \in \mathbb{R}^{d_{m} \times d_v}, W^O \in \mathbb{R}^{hd_v \times d_m}
    \label{eq1}
\end{split} 
\end{align}

We used $d_k = d_v = d_m/h = 64$ in this work. The $Attention$ function took a set of queries as a matrix $Q$, the keys matrix $K$, and the values matrix $V$. The output of the $Attention$ layer was:
\begin{align}\label{eq3}
    Attention(Q, K, V) = softmax \left(\frac{QK^T}{\sqrt{d_k}} \right)V.
\end{align}

All training and testing were conducted on a single NVIDIA RTX A6000 GPU, utilizing CUDA version V12.2.140. To further validate our model, we monitored metrics such as loss and accuracy during the training phase and utilized a validation dataset to assess performance periodically. 

\subsection{Model Evaluation}
We evaluated the predictive model's performance by computing the Pearson correlation coefficient $r$ between the prediction and the target. To do so, we first computed the correlation of each individual using (\ref{p_correlation_1}), where $a$ was the concatenated target actions, and $b$ was the concatenated model predictions. 

\subsection{Predictability as a Biomarker}
The predictive model we developed generates predicted future actions for each co-pilot based on the behavioral and physiological data of the other two co-pilots. These predictions are then correlated with the co-pilots' actual actions to compute a unique correlation score for each individual. We employ (\ref{p_correlation_1}) to calculate the holistic team biomarker, which averages the predictability scores across the three co-pilots. 

\subsection{Generalized Linear Mixed-effect Model} \phantomsection\label{GLMM}
As an extension to the generalized linear model (GLM), the linear predictors of the generalized linear mixed-effects model (GLMM) contained random effects in addition to the usual fixed effects~\cite{breslow1993approximate}. We used the GLMM in Python (statsmodels \cite{Skipper_statsmodels_Econometric_and_2010}) to investigate the relationship between varied variables with team difference considered as random-effect \cite{seabold2010statsmodels}. The final regression formula of each model was listed in supplementary materials in general form:
\begin{equation}
    y = X\beta+Z\mu + \varepsilon,
\end{equation}
where $y$ is the outcome variable. $X$ represents the predictor variables. $\beta$ is a column vector of the fixed-effects regression coefficients, and $Z$ is the design matrix for the random effects (the random complement to the fixed $X$). $\mu$ is a vector of the random effects (the random complement to the fixed $\beta$), and $\varepsilon$ is a column vector of the residuals. The supplementary materials list all the details, including all models' fixed and random effects.

\backmatter




\section{Acknowledgements}
This work was supported by funding from the Army Research Laboratory’s STRONG Program (W911NF-19-2-0139, W911NF-19-2-0135, W911NF-21-2-0125) the National Science Foundation (IIS-1816363, OIA-1934968) the Air Force Office of Scientific Research (FA9550-22-1-0337) and a Vannevar Bush Faculty Fellowship from the US Department of Defense (N00014-20-1-2027).
\\
\bibliography{sn-bibliography}

\end{document}